\documentclass[journal=apchd5,manuscript=article]{achemso}
\usepackage[utf8]{inputenc}
\usepackage{amsmath}
\usepackage{achemso}
\usepackage{siunitx}
\usepackage{amssymb}
\usepackage{verbatim}
\usepackage{xr}

\makeatletter

\title{Multiplication of the orbital angular momentum of phonon polaritons via sublinear dispersion}

\author{Andrea Mancini}
\email{A.Mancini@physik.uni-muenchen.de}
\affiliation[LMU]
{Chair in Hybrid Nanosystems, Nanoinstitute Munich, Faculty of Physics, Ludwig-Maxilimians-Universit\"at M\"unchen, 80539 M\"unchen, Germany}

\author{Lin Nan}
\affiliation[LMU]
{Chair in Hybrid Nanosystems, Nanoinstitute Munich, Faculty of Physics, Ludwig-Maxilimians-Universit\"at M\"unchen, 80539 M\"unchen, Germany}

\author{Rodrigo Bert\'e}
\affiliation[Brazil]
{Instituto de Física, Universidade Federal de Goiás, 74001-970 Goiânia-GO, Brazil}
\alsoaffiliation{Chair in Hybrid Nanosystems, Nanoinstitute Munich, Faculty of Physics, Ludwig-Maxilimians-Universit\"at M\"unchen, 80539 M\"unchen, Germany}

\author{Emiliano Cort\'es}
\affiliation[LMU]
{Chair in Hybrid Nanosystems, Nanoinstitute Munich, Faculty of Physics, Ludwig-Maxilimians-Universit\"at M\"unchen, 80539 M\"unchen, Germany}

\author{Haoran Ren}
\affiliation[Australia]
{School of Physics and Astronomy, Monash University, Clayton, Victoria 3800, Australia}

\author{Stefan A. Maier}
\affiliation[Australia]
{School of Physics and Astronomy, Monash University, Clayton, Victoria 3800, Australia}
\alsoaffiliation{Chair in Hybrid Nanosystems, Nanoinstitute Munich, Faculty of Physics, Ludwig-Maxilimians-Universit\"at M\"unchen, 80539 M\"unchen, Germany}
\alsoaffiliation{Department of Physics, Imperial College London, London SW7 2AZ, United Kingdom}

\newcommand*{\addFileDependency}[1]{
\typeout{(#1)}
\@addtofilelist{#1}
\IfFileExists{#1}{}{\typeout{No file #1.}}
}\makeatother

\newcommand*{\myexternaldocument}[1]{%
\externaldocument{#1}%
\addFileDependency{#1.tex}%
\addFileDependency{#1.aux}%
}

\myexternaldocument{Supplementary_Info}

\begin{document}

\begin{abstract}
Optical vortices (OVs) promise to greatly enhance optical information capacity via orbital angular momentum (OAM) multiplexing. The need for on-chip integration of OAM technologies has prompted research into subwavelength-confined polaritonic OVs. However, the topological order imprinted by the structure used for the transduction from free-space beams to surface polaritons is inherently fixed after fabrication. Here, we overcome this limitation via dispersion-driven topological charge multiplication. We switch the OV topological charge within a small $\sim 3 \%$ frequency range by leveraging the strong sublinear dispersion of low-loss surface phonon polaritons (SPhP) on silicon carbide membranes. Applying the Huygens principle we quantitatively evaluate the topological order of the experimental OVs detected by near-field imaging. We further explore the deuterogenic effect, which predicts the coexistence of multiple topological charges in higher-order polaritonic OVs. Our work demonstrates a viable method to manipulate the topological charge of polaritonic OVs, paving the way for the exploration of novel OAM-enabled light-matter interactions at mid-infrared frequencies.
\end{abstract}

\pagestyle{plain}

\section{Introduction}\label{sec1}

Vortices with defined topological features are an ubiquitous phenomenon in physics, appearing in different systems including superconductors \cite{abrikosov1957magnetic}, superfluids \cite{lounasmaa1999vortices}, exciton polaritons \cite{lagoudakis2008quantized, lerario2017room} and magnetic materials \cite{lin2014topological,chmiel2018observation}. Recently, propagating vortex beams characterized by twisted helical wavefronts carrying intrinsic orbital angular momentum (OAM) have been realized with electrons \cite{uchida2010generation, verbeeck2010production}, neutrons \cite{clark2015controlling, sarenac2022experimental} and helium atoms \cite{luski2021vortex}, providing a new emerging tool for the investigation of fundamental particles and interactions \cite{ivanov2022promises}. Optical vortices (OVs), the photon counterpart of such beams, constitute a theoretically unbounded set of orthogonal OAM modes, promising enhanced optical \cite{wang2012terabit, yan2014high, willner2015optical} and quantum \cite{mair2001entanglement, nagali2009quantum, fickler2014interface} information processing via OAM multiplexing. Pioneered by Allen \cite{allen1992orbital}, OVs are most usually generated by spiral phase plates \cite{yan2014high}, spatial light modulators (SLM) \cite{wang2012terabit}, and more recently with metasurfaces \cite{pu2015catenary, devlin2017arbitrary,dorrah2021structuring, ren2021nanophotonic, de2022radially, ahmed2022optical}. Regardless of the method, the minimum size of free-space OVs is fundamentally limited by the diffraction limit of light, while reduction of the OVs footprint is required for on-chip integration \cite{genevet2012holographic, ren2016chip,yue2018angular,ji2020photocurrent, ren2021orbital} and unveiling of new light-matter interaction regimes \cite{machado2018shaping, konzelmann2019interaction, ni2021gigantic}.

A route for realizing subwavelength OVs is via coupling optical beams into highly confined surface states such as surface plasmon polaritons (SPPs) \cite{maier2007plasmonics, kim2010synthesis, david2015nanoscale, david2016two, spektor2017revealing,yang2020deuterogenic,spektor2021orbital,frischwasser2021real, bai2022plasmonic,prinz2023orbital} and surface phonon polaritons (SPhPs) \cite{caldwell2015low, xiong2021polaritonic, wang2022spin} through spin-orbit interactions \cite{bliokh2015spin}. However, the topological charge offered by the surface structure used for launching polaritonic OVs is always a fixed quantity after device fabrication. The imprinted topological charge $L$  due to propagation phase-delay is proportional to the polariton momentum $L\propto k$, and can be then varied by multiplication of the wavevector by an integer factor (see Supplementary Note \ref{L_multiplication_expl}).

The required frequency variation to achieve a certain momentum shift is governed by the dispersion, a general property of waves governing the propagation and light-matter interactions of electromagnetic fields. As the dispersion of light in free-space is linear, doubling the momentum requires a $\sim100 \%$ frequency increase, making this approach impractical for most applications (Fig. \ref{FigIntro}a, blue line). In contrast, the required frequency variation for polaritonic OVs is significantly reduced due to their deeply sublinear dispersion (Fig. \ref{FigIntro}a, red curve). This small frequency change opens the possibility of flexible manipulation of the topological charge of polaritonic OVs. Even though the SPPs dispersion departs from the linear behavior, this occurs only in the region of high material losses \cite{maier2007plasmonics, khurgin2015deal}. SPhPs are characterized by low losses even in the deeply sublinear dispersion regime due to the reduced phonon-phonon scattering rate \cite{caldwell2015low}. As such, SPhPs vortices have recently been realized in a hexagonal boron nitride film on gold \cite{wang2022spin}, however the measured OVs near-field profiles were distorted due to tilted illumination, obscuring a clear evaluation of the topological order.

Here we demonstrate dispersion-driven OAM multiplication in highly confined SPhP vortices based on suspended \SI{100}{\nm} thick silicon carbide (SiC) membranes \cite{mancini2022near}. We realize the multiplication of polaritonic topological orders based on different surface structures with only a small $\sim 3\%$ variation in the excitation frequency (Fig. \ref{FigIntro}b). By introducing a set of reference functions accounting for material losses, we quantitatively determine the topological order of the experimental SPhP vortices mapped via a normal-incidence transmission-type near-field microscope (Fig. \ref{FigIntro}c). We precisely track the evolution of the SPhP OVs with high mode purity through a small frequency tuning. Furthermore, we investigate the deuterogenic effect predicting the coexistence of multiple OAM modes in surface structures producing higher order polaritonic OVs \cite{yang2020deuterogenic}. As a result, our demonstration presents a viable approach to flexible manipulation of the topological charge of polaritonic OVs, paving the way for the exploration of novel OAM-enabled light-matter interactions at the subwavelength scale.

\section{Results}

\begin{figure}[htp]
\centering
\includegraphics[width=1\textwidth]{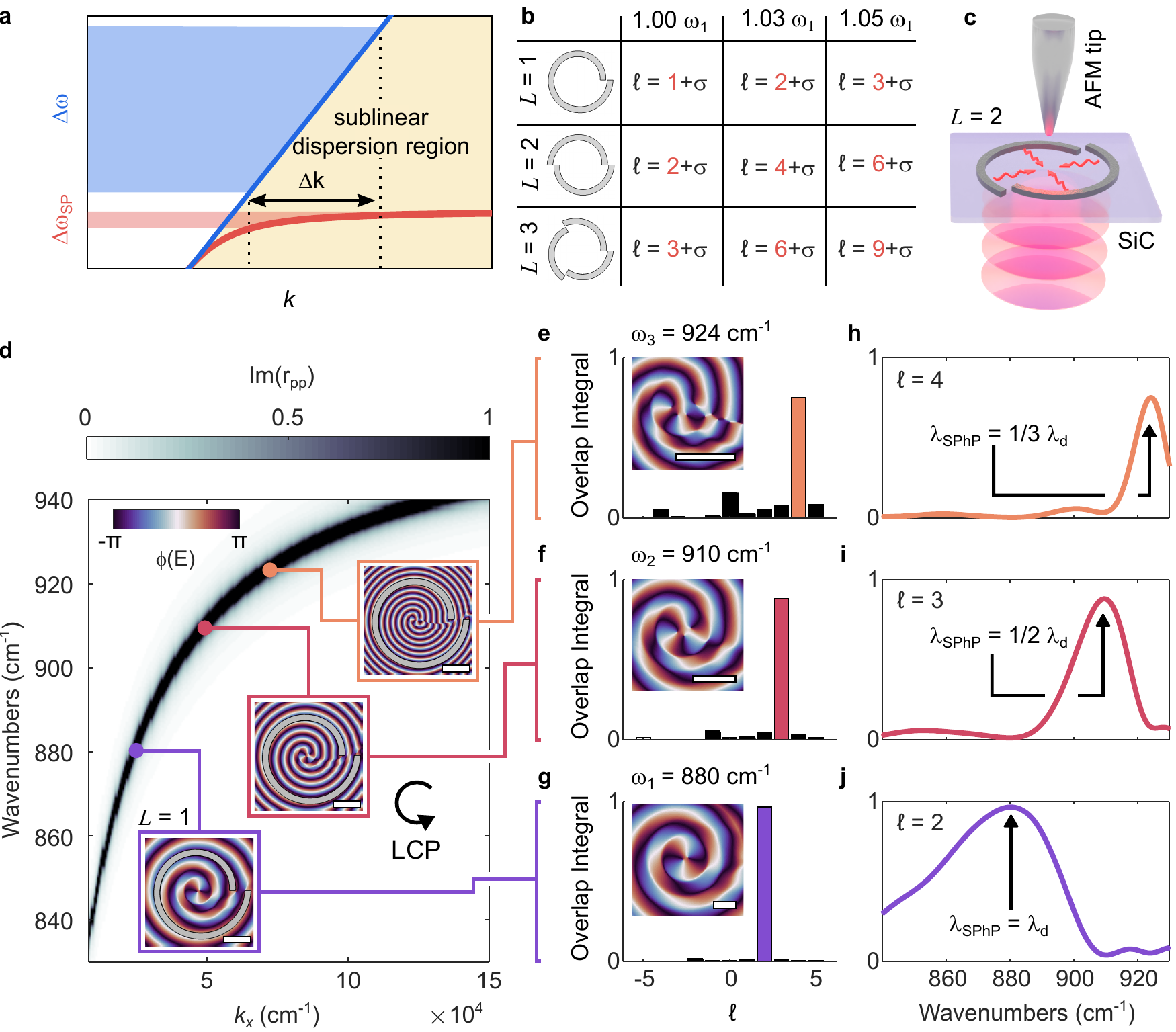}
\caption{\textbf{Topological vortex multiplication via sublinear SPhP dispersion}. \textbf{a} Dispersion-driven tuning for free-space and polaritonic OVs based on azimuthally-varying phase delay. The strong sublinear dispersion of surface polaritons results in a smaller frequency increase $\Delta \omega_{SP}$ for the same $\Delta k$ compared to the linear case $\Delta \omega$. \textbf{b} Changing the excitation frequency so that the momentum is increased by an integer number $nk$ leads to OAM multiplication. The three frequencies are chosen so that the polariton momentum is $k_1$, $2k_1$ and $3k_1$. $\sigma=\pm1$ is the spin state of the incident beam. \textbf{c} Sketch of the transmission sSNOM setup used to probe the near-field of the polaritonic OVs. \textbf{d} Calculated dispersion of the anti-symmetric mode of a \SI{100}{\nm} thick SiC membrane. As insets, simulated near-field phase maps for a SPhP vortex at three different frequencies launched by the same VG designed at $\omega = \SI{880}{\per \cm}$ with $L=1$ and excited by LCP light. Scale bar: $\SI{2}{\micro \meter}$. \textbf{e-g} Overlap integral at $\omega = \SI{924}{\per \cm}$, $\omega = \SI{910}{\per \cm}$ and $\omega = \SI{880}{\per \cm}$ used to evaluate the vortex purity. Insets: zoom of the maps shown in \textbf{d}. Scale bar: $\SI{1}{\micro \meter}$ \textbf{h-k} Overlap integral for orders $\ell =4$, $\ell =3$ and $\ell =2$ as a function of excitation frequency. $\lambda_d$ indicates the SPhP wavelength used for the VG design.}
\label{FigIntro}
\end{figure}

In polar thin films, hybridization between the individual SPhPs supported by the top and bottom interfaces occurs when the film thickness becomes comparable to the material skin depth \cite{mancini2022near}. In this regime splitting of the SPhP dispersion is observed, with the two modes characterized by symmetric and anti-symmetric out-of plane electric field distributions $E_z$ \cite{dionne2005planar}. In this paper we focus on the anti-symmetric mode as it features lower losses and longer propagation lengths (see Supplementary Note \ref{SPhP_film_dispersion}). The dispersion relation for the anti-symmetric mode of a \SI{100}{\nm} SiC membrane calculated through the transfer-matrix approach \cite{passler2017generalized} is shown in Fig. \ref{FigIntro}d. 

 We show the dispersion-driven multiplication of a SPhP OV topological order by simulating its out-of plane electric field component through the Huygens principle \cite{teperik2009huygens, tsesses2019spin} (see Supplementary Note \ref{Huygens_sim}). Multiple point sources are arranged in the shape of a vortex generator (VG) (see Supplementary Note \ref{Huygens_sim}) defined by the SPhP wavelength $\lambda_{SPhP}$, the topological order $L$ and its minimum radius $M\lambda_{SPhP}$. A VG with $L=1$ and $M=2$ designed at \SI{880}{\per \cm} (where the predicted SPhP wavelength is $\lambda_d = \SI{2.5}{\micro \meter}$) is used for the simulations shown in Fig. \ref{FigIntro}d. Only the phase of the calculated electric field is shown. We use the convention that anti-clockwise rotation is associated with a $+$, while clockwise rotation with a $-$ sign. The SPhP phase profile under circularly polarized excitation is shown in Fig. \ref{FigIntro}d at \SI{880}{\per \cm}, \SI{910}{\per \cm} and \SI{924}{\per \cm} where the SPhP wavelength is $\lambda_{SPhP}^{\SI{910}{\per \cm}} = \lambda_d/2$ and $\lambda_{SPhP}^{\SI{924}{\per \cm}} = \lambda_d/3$, corresponding to a multiplication of $k_{SPhP}$ by a factor of $2$ and $3$. The total OAM $\ell$ of the polaritonic OV at the design wavelength is given by $\ell = L + \sigma + l$, where $l$ and $\sigma=\pm1$ are the OAM and spin states of the incident beam. In the rest of the paper we only consider cases where the exciting beam carries no OAM, so that $l=0$. The SPhP vortex OAM is $\ell^{\SI{880}{\per \cm}} = 2$, $\ell^{\SI{910}{\per \cm}} = 3$ and $\ell^{\SI{924}{\per \cm}} = 4$, as the geometrical phase delay imprinted by the VG changes from $2\pi$ at \SI{880}{\per \cm} to $6\pi$ at \SI{924}{\per \cm}.

To quantitatively assess the purity of the obtained polaritonic OVs, an overlap integral between the simulated or experimental data with a set of reference functions is commonly used \cite{yang2020deuterogenic, de2022radially}. While for free-space OAM beams Laguerre-Gaussian (LG) modes are employed as reference, these are not suited to represent the characteristic spiral phase pattern appearing in polaritonic OVs, which is a result of material losses. We calculate from an integral approach the $E_z$ of a SPhP vortex at various $\ell$, constituting a set of suitable basis functions (see Supplementary Note \ref{Overlap_inbtegral_note}). The overlap integral can be computed as a function of both the topological order $\ell$ and of the frequency (see Supplementary Note \ref{Overlap_inbtegral_note}). In Fig. \ref{FigIntro}e-g the overlap spectra for the simulated maps reported in Fig. \ref{FigIntro}d are shown. The spectrum in Fig. \ref{FigIntro}g shows the high purity of the simulated $\ell = 2$ vortex. The purity decreases in  Fig.\ref{FigIntro}f, e, but the spectra clearly peak at $\ell = 3$ and $\ell = 4$, as expected. The overlap integral evaluated at fixed $\ell$ as a function of frequency is shown in Fig.\ref{FigIntro}h-j. A clear peak can be observed for $\ell=4$ at \SI{924}{\per \cm}, for $\ell=3$ at \SI{910}{\per \cm} and for $\ell=2$ at \SI{880}{\per \cm}. Intriguingly, these plots demonstrate that from theoretical simulations a single fixed VG can lead to the creation of vortices with different topological orders. By leveraging the strong sublinear SPhP dispersion, a slight tuning of the incident frequency is sufficient to multiply the polaritonic vortex order.

To experimentally demonstrate the creation of SPhP vortexes, we map the near-field pattern of the polaritonic OVs by employing a transmission scattering-scanning near-field optical microscope (sSNOM) as sketched in Fig. \ref{FigIntro}c. Transmission sSNOM has the advantage of normal incidence excitation and reduced tip-sample interaction, and has been the method of choice to investigate complex SPP patterns \cite{david2015nanoscale, david2016two,tsesses2018optical, tsesses2019spin} and resonating plasmonic nanoantennas \cite{schnell2009controlling, schnell2010phase, alonso2011real,bohn2015near}. We avoid using reflection sSNOM, where light is focused on the tip at a tilted angle, as this inevitably distorts the OVs pattern for multiple reasons. First, the VG design assumes excitation with a normal-incident plane wave; and second, the edge-launched SPhP fringes periodicity in reflection sSNOM depends on the relative angle between the incident beam k-vector and the edge in-plane normal direction \cite{kaltenecker2020mono,sternbach2020femtosecond,mancini2022near,luan2022imaging}, which varies along the VG geometry. In the experiments, right (RCP) or left (LCP) circularly polarized light impinges from below the sample at normal incidence. SPhP are launched by a thin \SI{20}{\nm} Chromium (Cr) VG fabricated by standard electron beam lithography on a SiC membrane. The near-field from the OVs is scattered by a metallic AFM tip, which oscillates at $\Omega\approx \SI{280}{\kilo \hertz}$ for background suppression obtained by demodulation of the signal at $n\Omega$ with $n>1$. The width of the VG arms is chosen to be \SI{2.5}{\micro \meter}, which we confirm is appropriate for efficient SPhP launching (see Supplementary Note \ref{ridge_thick}).

\begin{figure}[ht]
\centering
\includegraphics[width=1\textwidth]{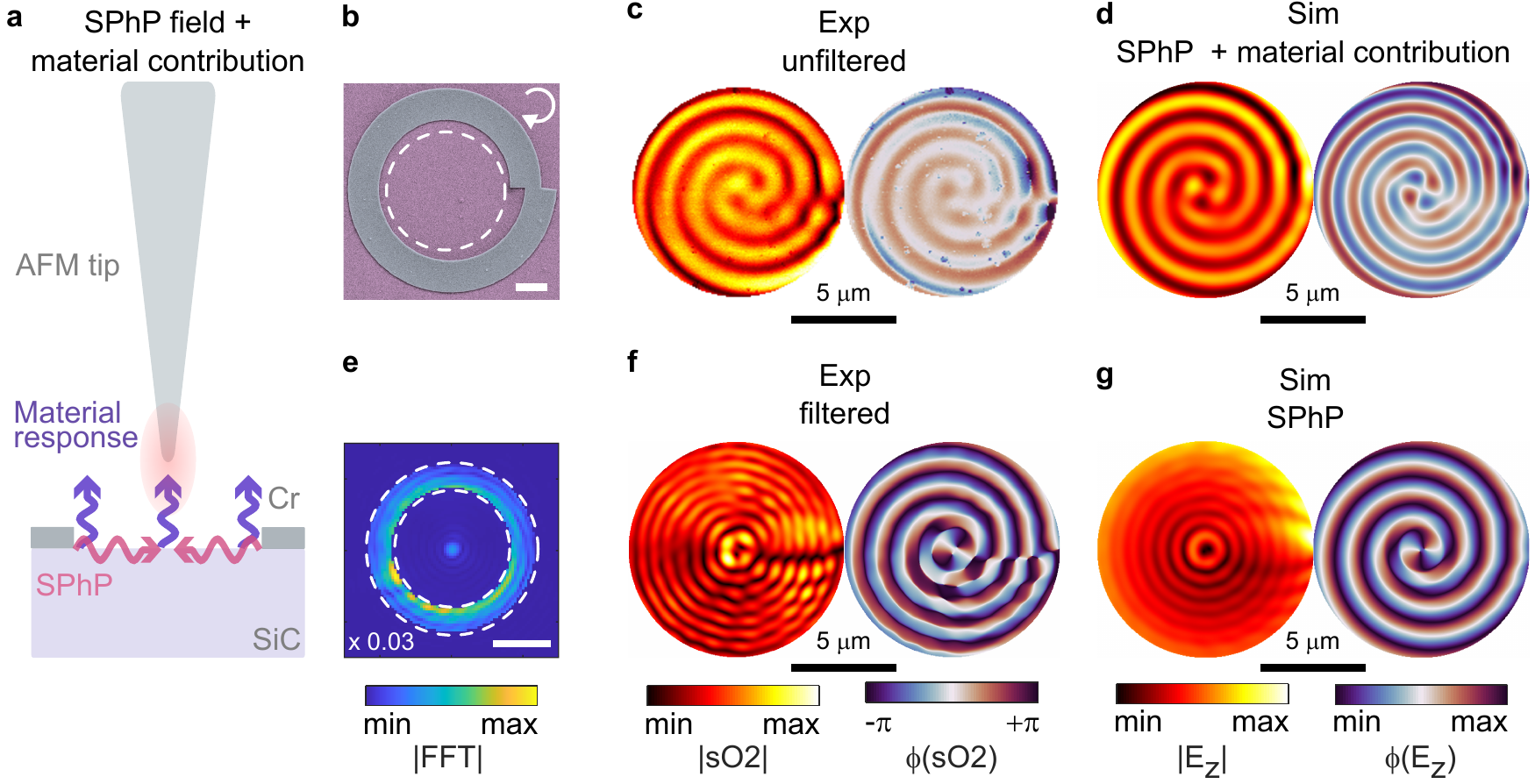}
\caption{\textbf{Experimental near-field mapping of a $\ell=2$ SPhP vortex.} \textbf{a} In sSNOM the pure SPhP field interferes with the bare material response as shown in the sketch. \textbf{b} False color SEM of a VG with $L=1$ excited by LCP light. Scale bar: \SI{2.5}{\micro \meter}. The dashed lines indicate the region where experimental data are shown.  \textbf{c} Experimental near-field amplitude and phase maps from the VG in \textbf{b} at $\omega = \SI{913}{\per \cm}$. \textbf{d} Simulated corresponding vortex profile, where a constant complex background is added to the SPhP $E_z$ field. \textbf{e} Removal of the material contrast through FFT filtering. Only the region between the dashed lines is kept. The rest of the map is shown multiplied by $0.03$. The bright spot at the center represents the position-independent material contrast. Scale bar: $10k_0$. \textbf{f} Experimental maps obtained by inverse FFT of the filtered data shown in \textbf{f}. \textbf{g} Simulated corresponding SPhP $E_z$ field.}
\label{FigExpDataord2}
\end{figure}

As sketched in Fig.\ref{FigExpDataord2}a, the sSNOM signal is composed by the interference between the SPhP field and the bare response of the material, also called material contrast \cite{tamagnone2018ultra}. As a consequence, the raw sSNOM signal is qualitatively different from the expected SPhP OV profile. In Fig. \ref{FigExpDataord2}b we report a SEM image of a VG with $M=3$ and $L = 1$, designed at $\omega = \SI{913}{\per \cm}$ and fabricated on top of a \SI{100}{\nm} SiC membrane. The VG is excited with LCP light at the design frequency, and the measurements are shown for the circular region indicated by the white dashed line. Experimental amplitude and phase maps demodulated at $2\Omega$ are shown in Fig. \ref{FigExpDataord2}c. The data are clipped and corrected as detailed in Supplementary Note \ref{exp_data_treatment}. The resulting vortex is deeply subwavelength as $\beta = \lambda_0/\lambda_{SPhP} = 7$ at this frequency. We demonstrate a maximum SPhP vortex confinement of $\beta=15.6$ at $\omega = \SI{930}{\per \cm}$, corresponding to $\lambda_{SPhP} = \SI{0.7}{\micro \meter}$ as reported in Supplementary Note \ref{SI_maximum_confinement}. The spiral pattern in the vortex amplitude can be attributed to the interference with the material background. At the same time the phase jumps along the SPhP propagation direction are expected to go from $-\pi$ to $\pi$, while we measure smaller oscillations. To reproduce the experimental maps, it is sufficient to add a constant complex material background $E_{mat}=A_{mat}\exp{(i\phi_{mat})}$ to the simulated SPhP field. By choosing appropriate values of $A_{mat}$ and $\phi_{mat}$, we obtain good agreement with the experimental maps as shown in Fig. \ref{FigExpDataord2}d. The position-independent material contribution can be removed by filtering the 2-dimensional fast fourier transform (FFT) of the experimental data as shown in Fig. \ref{FigExpDataord2}e. Only a region around the value of the predicted $k_{SPhP}$ is retained by multiplication with a mask that is $0$ outside the region highlighted by the white dashed lines in Fig. \ref{FigExpDataord2}e, and $1$ inside. In Fig. \ref{FigExpDataord2}e the region to be masked is multiplied by $0.03$ to show the dominant central peak associated with the material contrast. By inverse FFT, the bare SPhP contribution can be retrieved as shown in Fig. \ref{FigExpDataord2}f. The corresponding simulated SPhP field is shown in Fig. \ref{FigExpDataord2}g, demonstrating good agreement with the retrieved experimental maps. We evaluate the purity of the filtered experimental data in Fig. \ref{FigExpDataord2}f to be $>0.6$ (see Supplementary Note \ref{Overlap_inbtegral_note}), demonstrating the high accuracy of the generated and measured SPhP OV, allowing a clear evaluation the topological charge.

\begin{figure}[ht]
\centering
\includegraphics[width=1\textwidth]{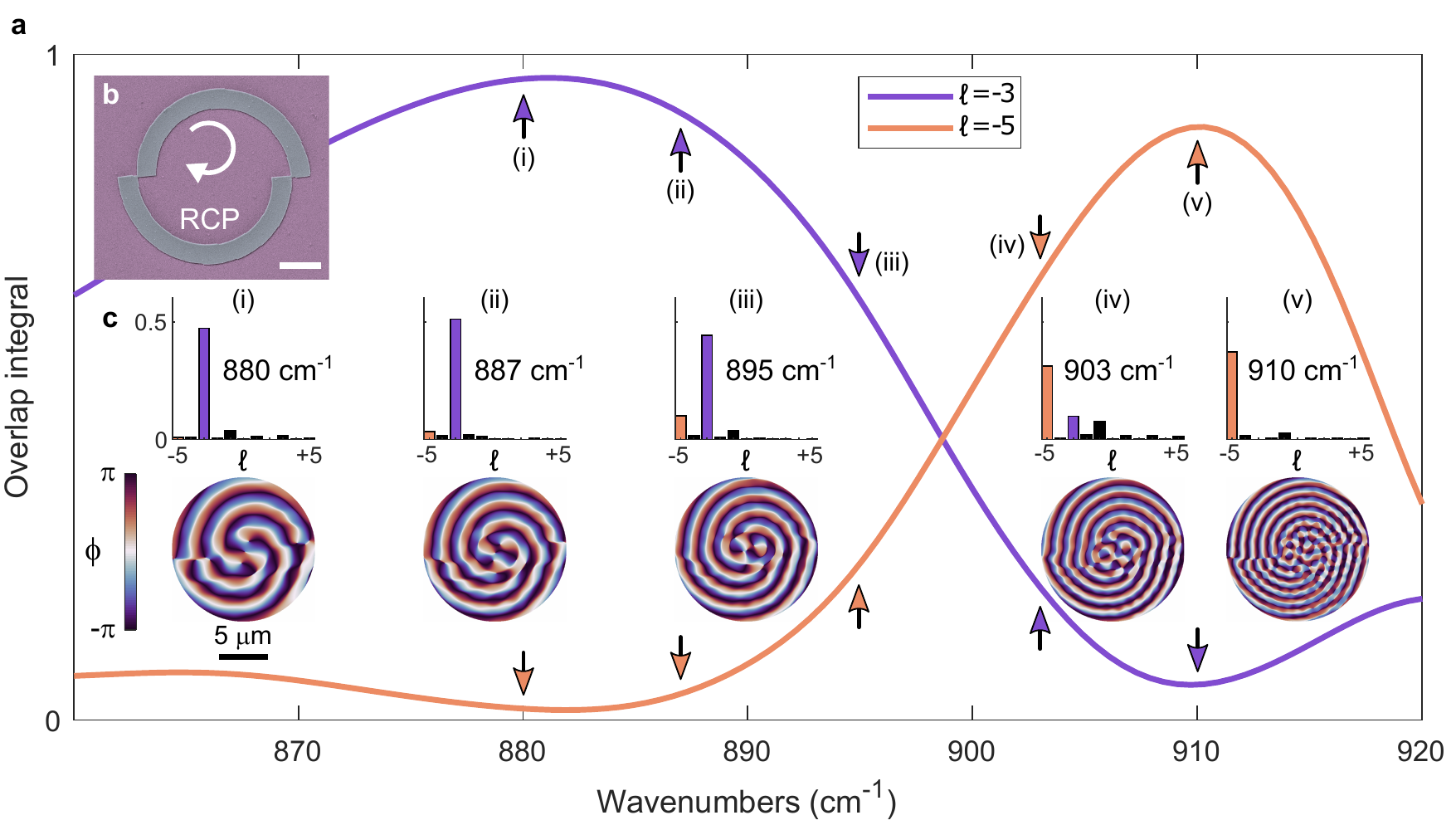}
\caption{\textbf{Experimental realization of topological order multiplication via SPhP dispersion}. \textbf{a} Simulated overlap integral as a function of frequency for a $L=-2$ VG excited with RCP light designed at $\omega=\SI{880}{\per \cm}$. Purple and orange lines indicate overlap with $\ell=-3$ and $\ell=-5$ modes respectively. \textbf{b} False color SEM image of the $L=-2$ VG used for the experiments. Scale bar: \SI{2.5}{\micro \meter}. \textbf{c} Top: overlap integral of the experimental data at different frequencies. The purple and orange bars correspond to $\ell=-3$ and $\ell=-5$ respectively. Bottom: corresponding FFT filtered experimental phase maps. Arrows and roman numbers indicate the position of the predicted overlap integral at the corresponding experimental frequencies.}
\label{Fig_tuning_dispersion_exp}
\end{figure}

To experimentally confirm the dispersion-driven OAM multiplication, we quantitatively assess the topological order of a SPhP vortex when changing the excitation frequency. In Fig. \ref{Fig_tuning_dispersion_exp}a, we report the simulated integral overlap for a VG designed at $\omega = \SI{880}{\per \cm}$ with $L=-2$ and $M=2$ excited by $\sigma = -1$ RCP light. We show the obtained spectra for $\ell=-3$ and $\ell=-5$ with purple and orange curves, respectively. A peak at the design frequency $\omega = \SI{880}{\per \cm}$ for $\ell =-3$ can be seen in Fig. \ref{Fig_tuning_dispersion_exp}a labeled as (i). The spectrum for $\ell =-5$ instead shows a peak at $\omega = \SI{910}{\per \cm}$ indicated by (v), where $\lambda_{SPhP} = \lambda_d/2$ and $\lambda_d$ is the SPhP wavelength at the design frequency. In Fig. \ref{Fig_tuning_dispersion_exp}b we report a false-color SEM image of the fabricated VG. We carry out experiments at five different frequencies, $\omega = \SI{880}{\per \cm}$, $\omega = \SI{887}{\per \cm}$, $\omega = \SI{895}{\per \cm}$, $\omega = \SI{903}{\per \cm}$ and $\omega = \SI{910}{\per \cm}$, labeled from (i) to (v) in Fig. \ref{Fig_tuning_dispersion_exp}. For each wavelength we report in Fig. \ref{Fig_tuning_dispersion_exp}c the FFT filtered phase map along with the experimental overlap integral. We indicate with purple bars the contribution at $\ell = -3$ and with orange bars the one at $\ell = -5$. The simulated values of the overlap integrals at the measured frequencies are highlighted by purple and orange arrows for $\ell = -3$ and $\ell = -5$, respectively. The experimental data are in excellent agreement with the simulations, as we can track the gradual change from $\ell = -3$ to $\ell = -5$ while also quantifying intermediate states given by a superposition of both vortex modes.

\begin{figure}[ht]
\centering
\includegraphics[width=1\textwidth]{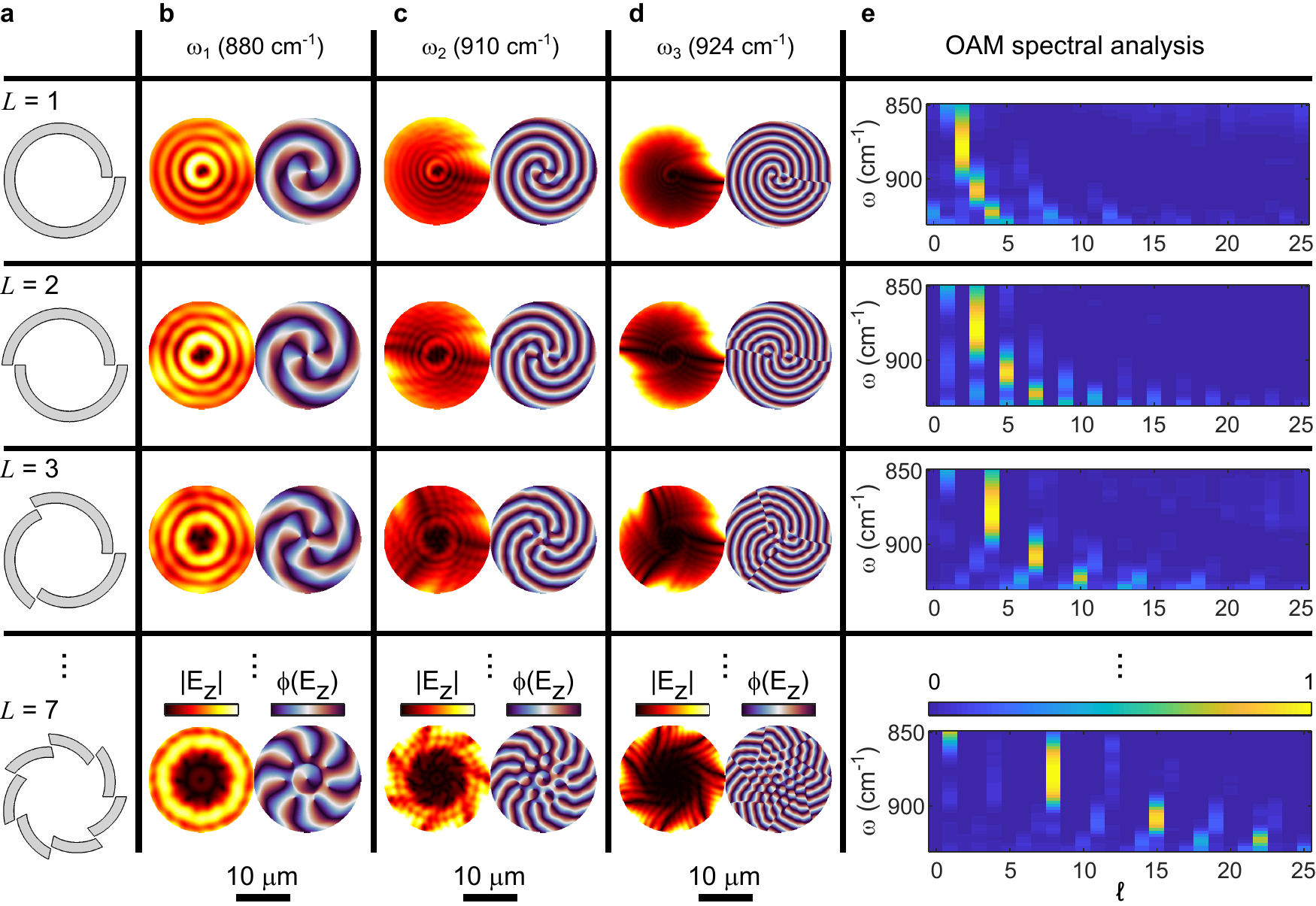}
\caption{\textbf{OAM multiplication for VG with different $\mathbf{L}$ under LCP excitation}. In \textbf{a} the shape of VGs with $L=1,2,3$ and $L=7$ is shown. Corresponding simulated amplitude and phase SPhP profiles at $\omega_1=\SI{880}{\per \cm}$ in \textbf{b}, at $\omega_2=\SI{910}{\per \cm}$ in \textbf{c} and at $\omega_3=\SI{924}{\per \cm}$ in \textbf{d}. For each row we calculate the overlap integral as a function of frequency and OAM order as shown in the maps of \textbf{e}. The maximum overlap is reached in all cases at frequencies $\omega_1, \omega_2$ and $\omega_3$. The $\ell$ spacing between the different frequencies is $L$, confirming the multiplication of the OAM order.}
\label{Fig_matrix}
\end{figure}

In Fig. \ref{Fig_tuning_dispersion_exp} we demonstrated experimentally the OAM multiplication for $L=-2$. We explore by simulations the same phenomena for different VGs excited by LCP light with $L=1,2,3$ and $L=7$ in Fig. \ref{Fig_matrix}. We report the data as a table, where the rows indicate the different investigated VGs. In of Fig. \ref{Fig_matrix}a we show sketches of the VGs. In Fig. \ref{Fig_matrix}b, c and d we report the simulated amplitude and phase maps at frequencies $\omega_1=\SI{880}{\per \cm}$, $\omega_2=\SI{910}{\per \cm}$ and $\omega_3=\SI{924}{\per \cm}$, at which the SPhP momentum is doubled and tripled with respect to the one at $\omega_1=\SI{880}{\per \cm}$. For each VG we calculate the overlap integral as a function of both frequency and topological order as shown in of Fig. \ref{Fig_matrix}e. The overlap integral peaks for all VGs at the three frequencies $\omega_1, \omega_2$ and $\omega_3$. The OAM multiplication operation can be observed by noticing that in Fig. \ref{Fig_matrix}e adjacent maxima in $\ell$ are spaced exactly by $L$. Therefore, the OAM multiplication can be applied in principle for any value of $L$.

\begin{figure}[htp]
\centering
\includegraphics[width=1\textwidth]{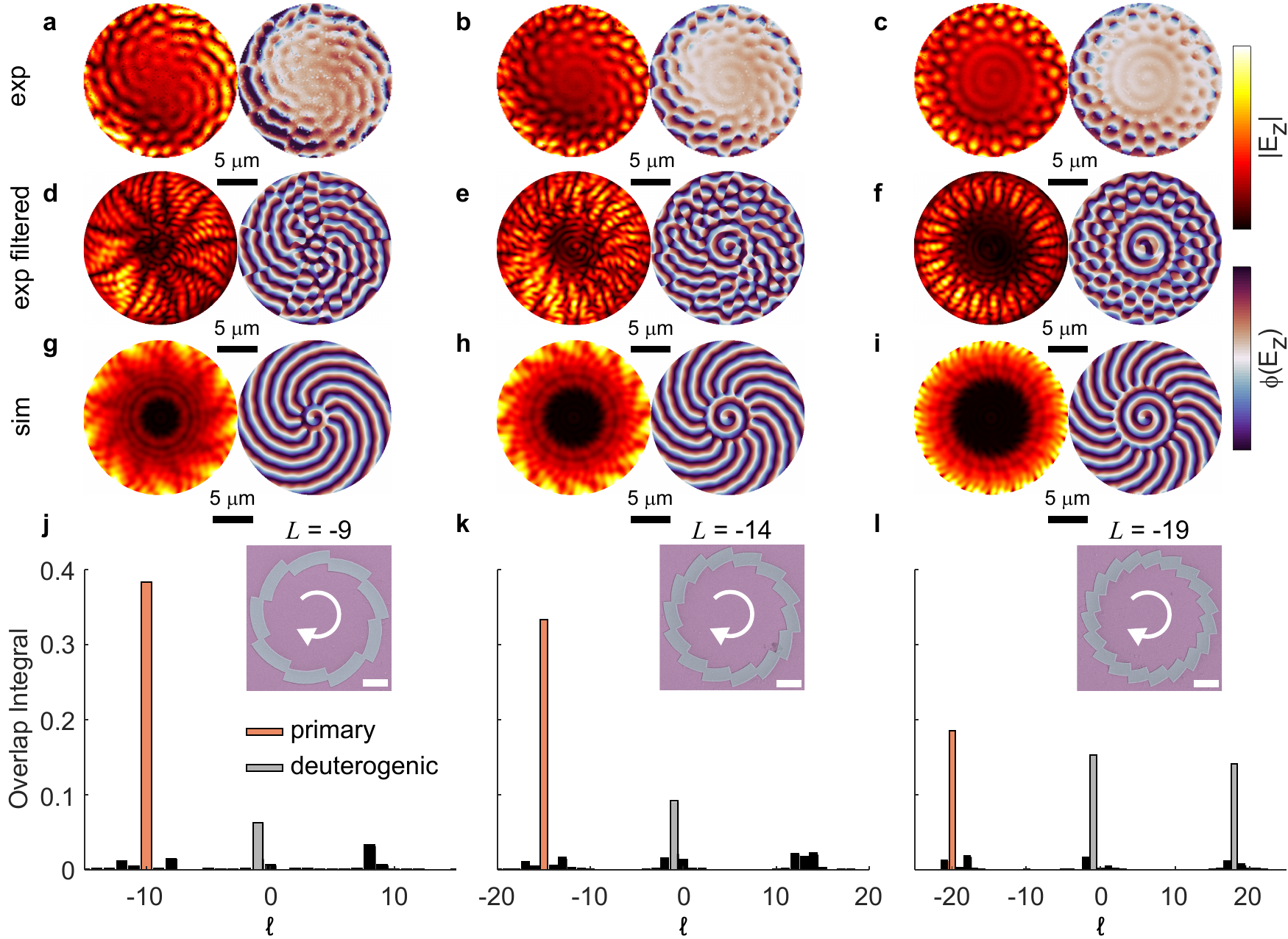}
\caption{\textbf{Deuterogenic effect in high order SPhP vortex at $\omega = \SI{900}{\per \cm}$}. Amplitude and phase transmission sSNOM maps at $\omega = \SI{900}{\per \cm}$ for VG of $L=-9$ \textbf{a}, $L=-14$ \textbf{b} and $L=-19$ \textbf{c} excited by RCP light. \textbf{d-f} Experimental maps after FFT filtering. \textbf{g-i} Corresponding simulated SPhP vortex. \textbf{j-l} Overlap integrals evaluated for the experimental data. Orange bars indicate the expected $\ell = -10$ in \textbf{j}, $\ell = -15$ in \textbf{k} and $\ell = -20$ in \textbf{l}. Gray bars indicated values of $\ell$ at which we detect the coexistence of additional modes predicted from the deuterogenic effect. Insets show false-color SEM of the fabricated VG used in the experiments. Scale bar: \SI{5}{\micro \meter}}
\label{Fig_high_order_vortex}
\end{figure}

In Fig. \ref{Fig_matrix} we showed that very high OAM orders can be obtained through the multiplication procedure, as for the case $L=7$ we obtain a maximum order of $\ell=22$ for $\omega=\SI{924}{\per \cm}$. We demonstrate that we can create and effectively measure these high order OVs. In particular we look into the coexistence of multiple OAM orders in polariton OVs, which has been recently shown to occur for SPPs \cite{yang2020deuterogenic}. This phenomena has been referred to as the "deuterogenic effect" to indicate the presence of secondary OVs as a consequence of the VG geometry used for the vortices creation. According to this principle, the total OAM order of a polariton OV is given by \cite{yang2020deuterogenic}:

\begin{equation}\label{deuterogenic}
   \ell = nL +\sigma + l  
\end{equation}

with $n=0,\pm1,\pm2,\ldots$ and $l$ and $\sigma=\pm1$ the OAM and spin states of the excitation beam. In Fig. \ref{Fig_high_order_vortex}a-c we report amplitude and phase near-field maps for vortices of order $\ell=-10$, $\ell=-15$ and $\ell=-20$ respectively. The measurements are performed at $\SI{900}{\per \cm}$ with RCP excitation. As the footprint of higher order vortices is larger, we use here $M=5$ for the definition of the VGs. In Fig. \ref{Fig_high_order_vortex}d-f we report filtered amplitude and phase maps, together with the corresponding simulations shown in Fig. \ref{Fig_high_order_vortex}g-i. We calculate the associated overlap integrals as displayed in Fig. \ref{Fig_high_order_vortex}j-l together with false-color SEM images of the Cr ridges used for SPhP launching. We observe that for higher $\ell$ the mode purity decreases. In Fig. \ref{Fig_high_order_vortex}j, k we detect the coexistence of modes $\ell=-10$ and $\ell=-15$ with the $\ell=-1$ order, as predicted from eq. \eqref{deuterogenic} for $n=0$ and $\sigma=-1$. The modes associated with the deuterogenic effect are indicated by gray bars. For the $\ell=-20$ vortex we can also observe a strong peak at order $\ell=18$, which is again expected from eq. \eqref{deuterogenic} for $n=-1$, $\sigma=-1$ and $L=-19$. While the deuterogenic effect for $n=0,1$ has been observed for SPP \cite{yang2020deuterogenic}, we report here the first experimental detection of a $n=-1$ contribution, indicating that further terms in eq. \eqref{deuterogenic} have to be considered to properly describe the generated polaritonic OV.  

Intuitively, the generation of the central $\ell=-1$ vortex can be associated with the SPhP launched by the corners of the VG arms, which overall constitute a circle of emitting dipoles. In this case there is no additional phase delay from the structure, and the vortex $\ell$ is equal to the sum of the SAM and OAM of the incident beam. This effect can be attenuated by reducing the number of arms of the VG and increasing the phase delay imprinted by each section of the VG (see Supplementary Note \ref{Minimization of the deuterogenic effect}). A balance between attenuation of the central lobe and the azimuthal distortion due to propagation losses has to be found depending on the application.

\section{Discussion}\label{sec3}

We leverage the strong sublinear SPhP dispersion to obtain vortices with multiplied topological orders by a small increase of the incident frequency of around $\sim 3\% $. This value can be further reduced by operating at higher frequencies where the group velocity is smaller. In comparison, an approximately $100\%$ increase in the operation frequency for free-space beams and SPPs \cite{bai2022generation} is required. As SPPs are sufficiently low-loss only when the dispersion is close to the light line, the smaller SPhP dissipation is a necessary ingredient to exploit in a practical way the dispersion-driven OAM multiplication mechanism. We experimentally demonstrated this tuning mechanism through near-field mapping of mid-IR SPhP vortex in SiC suspended membranes by transmission sSNOM. While reconfigurable polaritonic OVs can be always obtained by tuning the OAM order of the exciting beam, SLM are not readily available at mid-IR frequencies \cite{peng2015fast, fan2017graphene}, making our result particularly relevant for mid-IR technologies. Moreover, varying the incident beam OAM inevitably results in a reduced SPhP launching efficiency due to the diminished overlap between the surface structure and the high intensity part of the beam, which can be optimized only for a specific OAM value. The SiC membranes employed in this work are commercially available as millimeter scale chips and are obtained through CMOS-compatible fabrication, making this platform readily available for different applications. 

We find that the measured SPhP OVs agree with simulations, finding a mode purity $>0.6$ for a vortex with $\ell=2$. Thanks to the high quality of the measured SPhP vortices, we are able to investigate high OAM states with $\lvert \ell \rvert$ up to $20$. We also detect the coexistence of multiple OAM states for high order vortices as predicted by the deuterogenic effect \cite{yang2020deuterogenic}. The generation and precise control of highly pure SPhP vortices with high topological order paves the way for applications in structured thermal emission \cite{overvig2021thermal} and explorations of dipolar-forbidden transitions by highly confined electromagnetic fields carrying OAM \cite{machado2018shaping, konzelmann2019interaction}.

\section{Methods}\label{sec11}

\subsection*{Sample fabrication}

The suspended SiC films were purchased from Silson Ltd., UK. The spiral ridges were fabricated with standard electron-beam lithography followed by electron-beam deposition. First, the substrates were cleaned with O$_2$ plasma and spin-coated with \SI{120}{\nm} of 950K PMMA. Subsequently, the masks were fabricated under electron-beam exposure. After post-exposure development of the masks, the substrate was descummed with O$_2$ plasma prior to the electron-beam deposition of \SI{30}{\nm} of Cr. The structures were finally realized by lifting off the PMMA masks in acetone for several hours.

\subsection*{Transmission sSNOM}

Near-field spectra are obtained with a commercial sSNOM set-up (Neaspec) equipped with a pseudo-heterodyne interferometer to obtain amplitude and phase resolved images. The light source used in the experiments is an OPO laser (Stuttgart Instruments) powered by a pump laser at $\lambda=\SI{1035}{\nano \meter}$ with $\SI{40}{\mega \hertz}$ repetition rate and $\approx\SI{500}{\femto \second}$ pulses. The MIR output is obtained by DFG in a nonlinear crystal between the signal and idler outputs of the OPO and its frequency bandwidth reduced using a monochromator. A quarter-waveplate (Optogama) specifically designed to cover the SiC Reststrahlen band ($\approx800-\SI{950}{\per \cm}$ or $\approx 12.5-\SI{10.5}{\micro \meter}$) is used to convert the linear polarized output of the mid-infrared (IR) laser source (Stuttgart Instruments) to RCP and LCP light. The beam is loosely focused at normal incidence by a parabolic mirror positioned below the sample. The near-field signal is scattered by a metal-coated (Pt/Ir) atomic force microscope tip (Arrow-NCPt, Nanoworld) oscillating at a frequency $\Omega\approx\SI{280}{\kilo \hertz}$, and collected by a second off-axis parabolic mirror positioned above the sample. The tapping amplitude was set to $\approx\SI{80}{\nano\meter}$ and the signal demodulated at the second harmonic $2\Omega$ for background suppression. Before focusing, half of the light is redirected towards a pseudo-heterodyne interferometer used to retrieve both amplitude and phase of the signal. The light scattered by the tip is recombined with the interferometer reference arm by a second beam-splitter and directed towards a nitrogen-cooled mercury cadmium telluride infrared detector.

\subsection{Simulations}

The simulated vortex maps are obtained with a custom-written MATLAB code by using the Huygens principle (see Supplementary note \ref{Huygens_sim}). For each simulation at least $N=300$ individual sources are used. The dielectric function of SiC if modeled as follows:

\begin{equation}
    \varepsilon(\omega) = \varepsilon_{\infty} \left( 1+ \frac{\omega_{LO}^2 - \omega_{TO}^2}{\omega_{TO}^2 - \omega^2 - i\gamma \omega}\right)
\end{equation}

with $\omega_{TO} = \SI{797}{\per \cm}$, $\omega_{LO} = \SI{973}{\per \cm}$, $\varepsilon_{\infty}=6.6$ and $\gamma = \SI{6.6}{\per \cm}$. The value of $\gamma$ is somewhat higher than what recently estimated in SiC membranes \cite{mancini2022near}, but we use it as it better reproduces the experimental data. The SPhP wavevector in the SiC membranes is calculated numerically from the implicit equation for the anti-symmetric mode in polaritonic isotropic thin films \cite{mancini2022near}. 

\subsection{Acknowledgments}
S.A.M. acknowledges the Deutsche Forschungsgemeinschaft (MA 4699/7-1 and MA 4699/9-1), the Australian Research Council (DP220102152), and the Lee Lucas Chair in Physics. R.B. acknowledges the National Council for Scientific and Technological Development (CNPq, PDJ 2019—150393/2020-2). H.R. acknowledges the funding support from the Australian Research Council (DECRA Project DE220101085). This work was performed in part at the Melbourne Centre for Nanofabrication (MCN) in the Victorian Node of the Australian National Fabrication Facility (ANFF).

\bibliography{SiC_OAM.bib}


\end{document}